\documentclass{article}
\usepackage{dcase2024_techrep,amsmath,graphicx,url,times,booktabs, tabularx}
\usepackage{comment}

\usepackage{hhline}
\usepackage{hyperref}
\usepackage{xcolor}
\usepackage{cite}
\usepackage{multirow,url,hyperref}
\usepackage{siunitx}

\usepackage{pifont}
\newcommand{\cmark}{\ding{51}}%
\newcommand{\xmark}{\ding{55}}%
\usepackage{multirow}

\title{Improving Audio Spectrogram Transformers for Sound Event Detection through Multi-Stage Training}

\name{Florian Schmid$^1$, Paul Primus$^1$, Tobias Morocutti$^1$,  Jonathan Greif$^{1}$, Gerhard Widmer$^{1,2}$}
\address{$^1$Institute of Computational Perception (CP-JKU),$^2$LIT Artificial Intelligence Lab,\\          
        Johannes Kepler University Linz, Austria \\
        \{florian.schmid, paul.primus\}@jku.at\\ 
 }

\begin{document}

\ninept

\maketitle

\begin{sloppy}

\begin{abstract}
This technical report describes the CP-JKU team's submission for Task 4 \textit{Sound Event Detection with Heterogeneous Training Datasets and Potentially Missing Labels} of the DCASE 24 Challenge. We fine-tune three large Audio Spectrogram Transformers, PaSST, BEATs, and ATST, on the joint DESED and MAESTRO datasets in a two-stage training procedure. The first stage closely matches the baseline system setup and trains a CRNN model while keeping the large pre-trained transformer model frozen. In the second stage, both CRNN and transformer are fine-tuned using heavily weighted self-supervised losses. After the second stage, we compute strong pseudo-labels for all audio clips in the training set using an ensemble of all three fine-tuned transformers. Then, in a second iteration, we repeat the two-stage training process and include a distillation loss based on the pseudo-labels, boosting single-model performance substantially. Additionally, we pre-train PaSST and ATST on the subset of AudioSet that comes with strong temporal labels, before fine-tuning them on the Task 4 datasets\footnote{Code: \url{https://github.com/CPJKU/cpjku_dcase24}}.

\end{abstract}

\begin{keywords}
DCASE Challenge, Sound Event Detection, ATST, BEATs, PaSST, DESED, MAESTRO, pseudo-labels
\end{keywords}


\section{Introduction}
\label{sec:intro}

The task of Sound Event Detection (SED) is to recognize and classify specific sound events in audio signals, including the temporal location of the events. Developing reliable SED systems allows their use in important real-world applications, such as security and surveillance~\cite{radhakrishnan2005audio}, smart homes~\cite{debes2016monitoring}, or health monitoring~\cite{zigel2009method}. A main driver of research in this field is the annual DCASE Challenge, with Task 4 specifically tackling Sound Event Detection. This technical report describes the CP-JKU team's submission to DCASE Challenge 2024 Task 4: \textit{Sound Event Detection with Heterogeneous Training Datasets and Potentially Missing Labels}~\cite{cornell2024dcase}.

State-of-the-art SED systems are based on deep learning approaches, requiring a substantial amount of annotated data. Their performance is mainly limited by the acute lack of strongly annotated sound event datasets~\cite{Martinmorato2023maestro}. Hence, previous editions of Task 4 focused on learning from weakly labeled data~\cite{kong2020sound}, semi-supervised learning strategies~\cite{park2021self}, and utilizing synthetic strongly labeled data~\cite{Turpault2019DCASE} in an attempt to develop systems that perform well on real-world strongly labeled sound clips. While Task 4 has been based on the DESED dataset~\cite{Turpault2019DCASE} in previous years, the key novelty of the 2024 edition is a unified setup including a second dataset, MAESTRO Real~\cite{Martinmorato2023maestro}. As domain identification is prohibited, the goal is to develop a single system that can handle both datasets despite crucial differences, such as labels with different temporal granularity and potentially missing labels. In fact, because of the lack of strongly annotated, high-quality real-world data, the hope is that learning from two datasets in parallel has a synergetic effect and eventually increases the performance on both datasets, as demonstrated for the baseline system~\cite{cornell2024dcase}. 

The main contributions of this work can be summarized in the following points: \textbf{(1)} We demonstrate that multiple different pre-trained transformer models (ATST~\cite{li2022atst}, PaSST~\cite{koutini22passt}, and BEATs~\cite{chen2022beats}) can be fine-tuned on the Task 4 datasets to achieve high performance. \textbf{(2)}~Pre-training on the temporally-strong annotated portion of AudioSet~\cite{gemmeke17audioset} (AudioSet strong) can improve performance for the frame-wise pre-trained model ATST and is necessary for the clip-wise pre-trained PaSST to obtain high-quality frame-wise predictions. \textbf{(3)} Combining fine-tuned ATST, PaSST, and BEATs models leads to a diverse ensemble that can be used to create high-quality pseudo-labels. \textbf{(4)} Using the computed pseudo-labels in a second training iteration dramatically improves single-model performance, leading to a relative increase of 25.6\% in terms of polyphonic sound detection score~\cite{ebbers2022threshold, bilen2020framework_psds1} (PSDS1) on DESED and 2.7\% in terms of segment-based mean partial area under the ROC curve (mpAUC) on MAESTRO compared to the baseline system. On DESED, we achieve a new state-of-the-art performance on the public evaluation set, increasing the single-system performance from .686~\cite{ebbers2022threshold} to .692 in terms of PSDS1.

\vspace{-5pt}
\section{Datasets}
\label{sec:datasets}

The development set is composed of two datasets: DESED~\cite{Turpault2019DCASE} and MAESTRO Real~\cite{Martinmorato2023maestro}. For common processing, all audio in the training set is converted to clips of 10 seconds in length. For the MAESTRO dataset, we strictly follow the train-test-validation split established by the baseline system~\cite{cornell2024dcase}. As for DESED, we use the following subsets:
\begin{itemize}
    \item Weakly labeled: clip-wise labels, 1,267 / 158 for train. / valid.
    \item Unlabeled: 13,057 unlabeled clips
    \item Synthetic strong: 10,000 / 2,500 strongly labeled synthetic clips for train. / valid.
    \item AudioSet strong: 3,435 strongly labeled real clips
    \item External strong: 6,426 / 957 additional strongly labeled real clips for train. / valid. from AudioSet strong as used in~\cite{xiao2023sound}
    \item Test: 1,168 strongly labeled real clips as in baseline setup~\cite{cornell2024dcase}
\end{itemize}

We noticed that the provided AudioSet strong subset overlaps with weakly labeled, unlabeled, and test sets. Therefore, we remove 1,355 files from the unlabeled set and 153 files from the weakly labeled set to avoid oversampling individual audio clips. Additionally, we remove the 35 files that overlap with the test set from the AudioSet strong set to obtain a more accurate estimate of the generalization performance.

\vspace{-5pt}
\subsection{Cross mapping sound event classes}

To exploit the fact that the DESED and MAESTRO classes are not fully disjoint but partly represent the same concepts, the baseline system introduces class mappings. For example, when the classes \textit{people talking}, \textit{children voices}, or \textit{announcement} are active in a MAESTRO clip, the corresponding DESED class \textit{Speech} is set to the same confidence value. 

In addition, we also include a mapping from MAESTRO classes to DESED clips. Specifically, we set the values of the MAESTRO classes \textit{cutlery and dishes} and \textit{people talking} to 1 if the DESED classes \textit{Dishes} and \textit{Speech} are present. This is also performed for weak class labels.

\vspace{-5pt}
\section{System Architecture}
\label{sec:arch}

Figure~\ref{fig:overview_sys} depicts an overview of our SED system. The system is very similar to the baseline~\cite{cornell2024dcase}. However, besides BEATs~\cite{chen2022beats}, we experiment with two additional Audio Spectrogram Transformers, ATST~\cite{li2022atst} and PaSST~\cite{koutini22passt}. Section~\ref{subsec:fpasst} introduces modifications to the PaSST architecture for allowing high-quality frame-wise predictions, and in Section~\ref{subsec:as_pre-training}, we describe pre-training of PaSST and ATST on AudioSet strong. In addition to adaptive average pooling, we experiment with linear and nearest-exact interpolation to align transformer and CNN sequence lengths. The BiGRU block consists of two bidirectional GRU layers with a dimension of 256.

\begin{figure}[t!]
\centering
{\includegraphics[width=\linewidth]{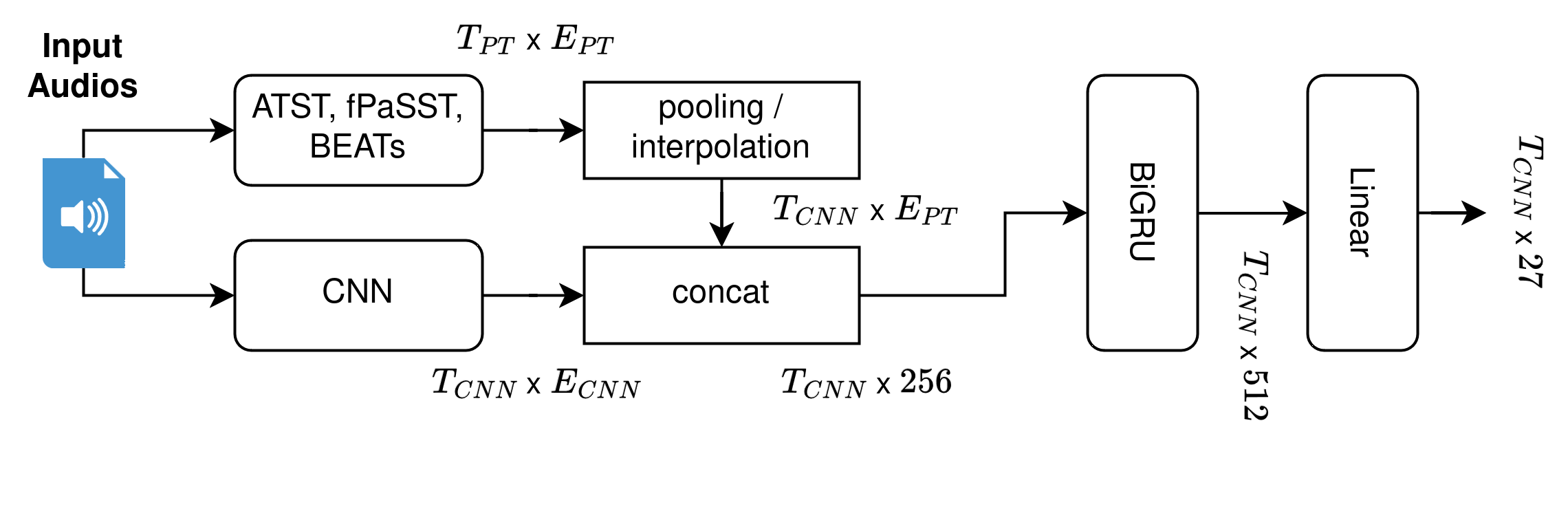}}
\vspace{-30pt}
\caption{Overview of System Architecture.}
\vspace{-10pt}
\label{fig:overview_sys}
\end{figure}

\vspace{-5pt}
\subsection{ATST-Frame}

ATST-Frame \cite{li2022atstf}(denoted only ATST in the following) was specifically designed to produce a sequence of frame-level audio embeddings instead of a single global clip-level representation and is thus particularly suited for SED. The architecture of ATST is based on that of the Audio Spectrogram Transformer (AST) \cite{gongast} and it is trained in a self-supervised manner via masked spectrogram modeling in a student-teacher scheme on AudioSet. In our experiments, we use a checkpoint of ATST that is further fine-tuned on the weak labels of AudioSet.

\subsection{fPaSST}
\label{subsec:fpasst}

The Patchout faSt Spectrogram Transformer (PaSST) \cite{koutini22passt} is an improved version of the original AST \cite{gongast} that shortens the training time and improves the performance via patchout regularization. PaSST uses global classification tokens, which are ideal for tagging and classification tasks but are not designed for inferring the precise temporal occurrence of acoustic events. We thus adopt the architecture of PaSST to return frame-level predictions and call the resulting model Frame-PaSST (fPaSST). fPaSST uses three input convolutions to convert the input spectrogram to a tensor of size $16 \times 128 \times 250$ (channel $\times$ frequency $\times$ time). The result is then converted to $250$ $768$-dimensional tokens via another convolution with kernel size $128 \times 2$. We modify the positional encoding accordingly and initiate all parameters except the first three convolutions with parameters taken from a vision transformer. We pre-trained the resulting model on the weakly annotated AudioSet using Knowledge Distillation as described in~\cite{schmid2023efficient}, obtaining a mAP of 0.484.

\vspace{-5pt}
\subsection{BEATs}
Likewise, BEATs \cite{chen2022beats} is also based on the AST \cite{gongast} architecture; it takes rectangular spectrogram patches as input and returns one embedding vector for each, making it suitable for SED. The model was trained in an iterative, self-supervised procedure where the BEATs encoder learned representations from a frozen audio-tokenizer model that was itself learned from the BEATs encoder after every iteration. In our experiments, we rely on the checkpoint of BEATs after the third iteration, where both the tokenizer and the encoder were fine-tuned on the weak labels of AudioSet.

\vspace{-5pt}
\section{Training Pipeline}
\label{sec:training_setup}

In this section, we describe the pre-training routine on AudioSet strong and how the pre-trained models are fine-tuned on the Task 4 datasets in a multi-iteration, multi-stage training procedure. An overview of the full training pipeline is shown in Figure~\ref{fig:overview_train}. The multi-stage setup closely follows the strategy outlined in~\cite{shao2024fine}, and a multi-iteration setup with learning from pseudo-labels showed to improve performance drastically in~\cite{ebbers2022pre} and~\cite{kim2023semi}. In the following, we abbreviate Iteration \{1,2\} and Stage \{1,2\} as I\{1,2\} and S\{1,2\}, respectively.

\begin{figure}[t!]
\centering
{\includegraphics[width=.8\linewidth]{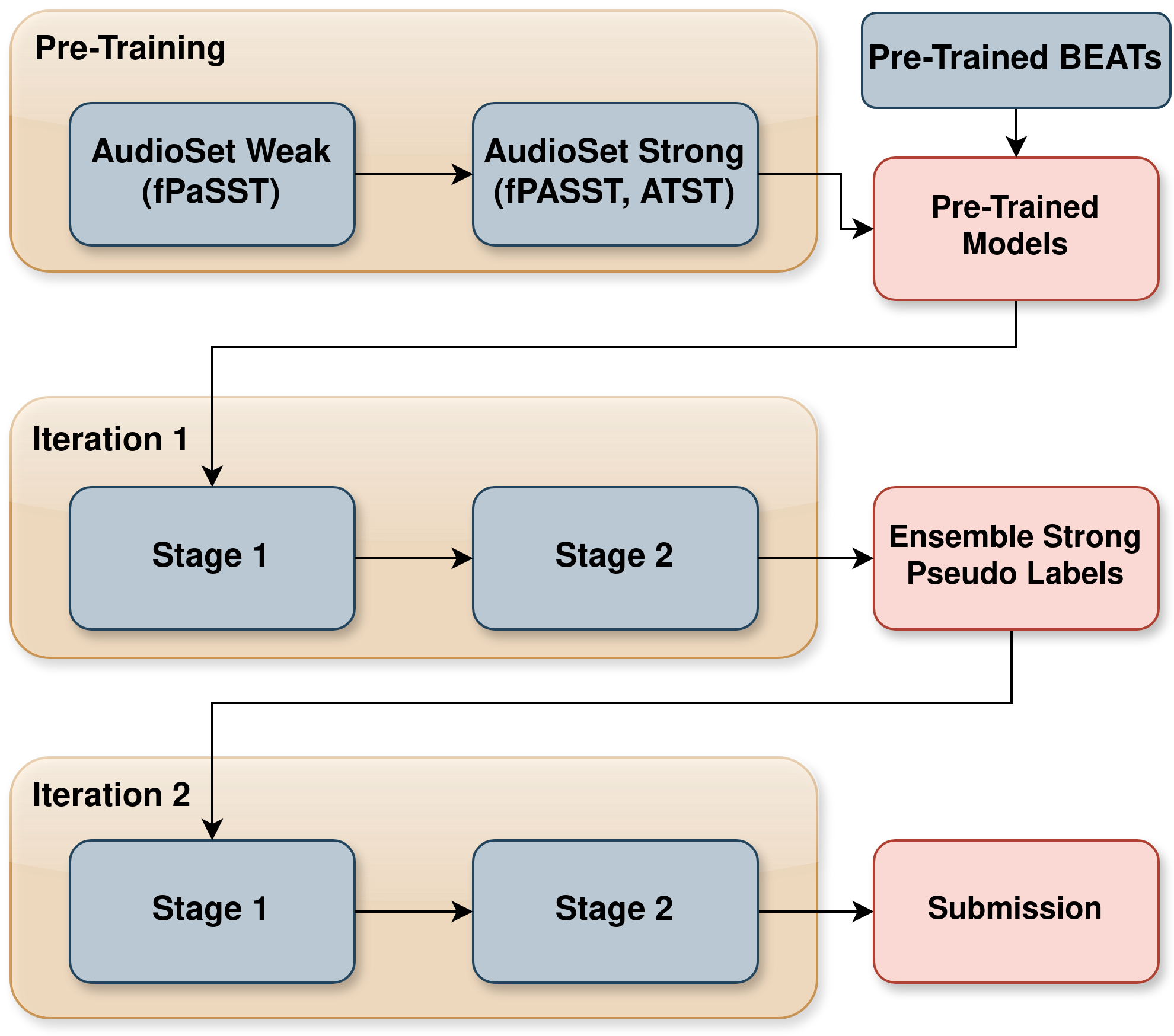}}
\caption{Overview of Training Pipeline.}
\vspace{-10pt}
\label{fig:overview_train}
\end{figure}

\vspace{-5pt}
\subsection{Pre-Training on AudioSet strong}
\label{subsec:as_pre-training}
We hypothesize that the audio embedding models would benefit from additional pre-training on a large dataset strongly annotated for various acoustic events. To this end, we add a BiGRU block with 1024 units that processes the output of fPaSST or ATST. We train both models for 10 epochs on AudioSet strong \cite{hersheyasstorng}, a subset of AudioSet that holds around 86.000 strongly labeled examples with annotations for 456 event classes. The learning rate for the pre-trained ATST and PaSST encoders is linearly increased from 0 to 4e-5, while the learning rate for the uninitialized BiGRU block is linearly decayed from 1e-3 to 4e-5 in the first four epochs. We select the checkpoint with the highest PSDS1 score on the AudioSet strong validation set for Task 4 downstream training.

\vspace{-5pt}
\subsection{Multi-Stage Training}

I1 and I2 are both split into two training stages. In S1, the CNN and BiGRU are trained from scratch while the large transformer model is kept frozen. This setup corresponds to the training of the baseline system with slightly different hyperparameters and additional data augmentations. 

In S2, the CNN and BiGRU are initialized with pre-trained weights from S1, and the transformer model is fine-tuned. As the system already performs well in its initial state, the transformer can rely on high-quality self-supervised loss computed on the larger unlabeled set. Aligned with~\cite{shao2024fine}, in S2, we compute interpolation consistency loss~\cite{zheng2021improved} in addition to the mean teacher loss.

In both stages, we choose the best model based on the validation metrics of the student. Specifically, we compute the sum of PSDS1 on the strongly labeled synthetic data, PSDS1 on external strongly labeled real data, and mpAUC on the MEASTRO validation set.

\vspace{-5pt}
\subsection{Multi-Iteration Training}

After completing I1, we build an ensemble (see \textit{Ensemble Stage 2} in Table~\ref{tab:results}) of multiple ATST, fPaSST, and BEATs models. This ensemble is used to compute strong pseudo-labels for all audio clips in the training set by averaging the frame-wise logits of the individual models. In S1 of I2, we then compute BCE between the model's predictions and the pseudo-labels as an additional loss term. We found that BCE is superior to MSE, and interestingly, using the pseudo-label loss only helps in S1 of I2. We hypothesize that the CRNN picks up relevant information from the pseudo-labels in S1 and transfers it to the transformer model via the high self-supervised loss weights in S2 of I2.

\vspace{-5pt}
\section{Experimental Setup}
\label{sec:exp_setup}

\vspace{-5pt}
\subsection{Audio Pre-processing}

For all models, we resample audio clips at 16 kHz. For the CNN, we match the baseline settings and compute mel spectrograms with 128 mel bins using a window length of 128 ms and a hop size of 16 ms. For ATST, fPaSST, and BEATs, we match the pre-training setup and compute mel spectrograms with 64, 128, and 128 mel bins, respectively. All transformers use a hop size of 10 ms; the window size of fPaSST and BEATs is set to 25 ms; and for ATST, it is 64 ms.

\vspace{-5pt}
\subsection{Data Augmentation}

\begin{table}[t]
\begin{center}
\begin{tabular}{@{}lccc@{}}
\toprule
\textit{Aug. Method} & \textbf{Target} & \textbf{HP} & \textbf{Pipeline} \\ \midrule
DIR~\cite{morocutti2023device} & All & $p$=0.5 & I\{1,2\}.S2 \\ 
Wavmix~\cite{zhang2017mixup} & Str. & $p$=0.5,$\alpha$=0.2 & I\{1,2\}.S\{1,2\}  \\ \midrule
Freq-MixStyle~\cite{schmid2022knowledge} & All & $p$=0.5,$\alpha$=0.3 & I1.S\{1,2\},I2.S2\\ 
Mixup~\cite{zhang2017mixup}  & All & $p$=0.5,$\alpha$=0.2 & I\{1,2\}.S\{1,2\} \\
Time-Masking & DES. Str. & s=[0.05,0.3] & I\{1,2\}.S2 \\ 
FilterAugment~\cite{nam2022filteraugment} & All & linear,p=0.8 & I1.S\{1,2\},I2.S2    \\ 
Freq-Warping~\cite{li2022atst} & All & p=0.5 & I\{1,2\}.S2   \\  \bottomrule
\end{tabular}
\caption{The table lists data augmentation methods, the data subset they are applied to (\textbf{Target}), hyperparameters (\textbf{HP}), and the respective iteration and stage they are used in (\textbf{Pipeline}). \textit{p} is the probability for applying the augmentation method; $\alpha$ parameterizes Beta distributions; and \textit{Str.} refers to strongly annotated audio clips.}
\label{tab:aug}
\end{center}
\vspace{-20pt}
\end{table}

Table~\ref{tab:aug} presents in detail all the data augmentation methods we use in our training pipeline. In contrast to the baseline, we apply Cross-Dataset Mixup and Cross-Dataset Freq-MixStyle. That is, we mix audio clips from MAESTRO and DESED instead of keeping them separate. In the case of Mixup, we modify the class mask and allow the loss to be calculated for all active classes, irrespective of the audio clip's dataset origin. For Wavmix and Mixup, we mix the pseudo-labels accordingly.

\vspace{-5pt}
\subsection{Data Sampling and Optimization}

We summarize the training data in five subsets: MAESTRO, DESED real strong, synth. strong, weakly annotated, and unlabeled. In S1, we draw batches of (12, 10, 10, 20, 20) samples, and in S2, we draw batches of (56, 40, 40, 72, 72) samples from these datasets. The model needs to optimize six losses in parallel: MAESTRO strong, DESED real strong, synth. strong, weak, self-supervised loss, and pseudo-label loss. Besides the MSE loss used for the self-supervised loss, the BCE loss is computed for all others. We compute a weighted sum of all losses and tune the individual weights for all iterations and stages. In contrast to the baseline, we also compute the self-supervised loss on MAESTRO clips. 

We use the AdamW~\cite{loshchilov2017decoupled} optimizer with weight decays of 1e-2 and 1e-3 in S1 and S2, respectively. Learning rates are listed in Table~\ref{tab:results}.

\begin{table*}[h!]
\centering
\vspace{-8pt}
\begin{tabular}{@{}l|l|l|ccccc|ccc@{}}
\toprule
    &   & \textbf{Model}  & \textbf{lr\_cnn} & \textbf{lr\_rnn} & \textbf{lr\_tf} & \textbf{lr\_dec} & \textbf{Seq.} & \textbf{mpAUC} & \textbf{PSDS1}  & \textbf{Rank Score}\\ 
     
      \midrule
\multirow{ 6}{*}{Iteration 1} & \multirow{ 3}{*}{Stage 1} & ATST  & 1e-3 & 1e-3 & - & - & int. lin. & 0.702 $\pm$ 0.008 & 0.493 $\pm$ 0.012 & 1.195 $\pm$ 0.012 \\
 & & fPaSST  & 1e-3 & 1e-3 & - & - & int. nearest & 0.709 $\pm$ 0.021 & 0.502 $\pm$ 0.010 & 1.212 $\pm$ 0.027 \\
 &  & BEATs & 1e-3 & 1e-3 & - & - & int. nearest & 0.719 $\pm$ 0.004 & 0.509 $\pm$ 0.003 & \textbf{1.228 $\pm$ 0.006} \\ \cline{2-11}
& \multirow{ 3}{*}{Stage 2} & ATST  & 1e-4 & 1e-3 & 1e-4 & 0.5  & int. nearest & 0.739 $\pm$ 0.017 & 0.520 $\pm$ 0.005 & \textbf{1.259} $\pm$ 0.020 \\
&  & fPaSST & 1e-4 & 1e-3 & 1e-4 & 1 & int. nearest & 0.726 $\pm$ 0.021 & 0.514 $\pm$ 0.008 & 1.24 $\pm$ 0.027 \\
&  & BEATs & 1e-4 & 1e-3 & 1e-4 & 1 & int. lin. & 0.713 $\pm$ 0.002 & 0.539 $\pm$ 0.004 & 1.252 $\pm$ 0.003 \\ \cline{2-11}
& \multicolumn{2}{c}{Ensemble Stage 2} & - & - & - & - & mix & 0.735 & 0.569 &  1.303 \\\hline
\multirow{ 6}{*}{Iteration 2} & \multirow{ 3}{*}{Stage 1} & ATST  & 5e-4 & 5e-4 & -  & - & avg. pool & 0.741 $\pm$ 0.017 & 0.536 $\pm$ 0.006 & \textbf{1.277 $\pm$ 0.012} \\
&  & fPaSST & 5e-4 & 5e-4 & - & - & int. nearest & 0.722 $\pm$ 0.011 & 0.526 $\pm$ 0.004 & 1.248 $\pm$ 0.012 \\
&   & BEATs & 5e-4 & 5e-4 & - & - & int. nearest & 0.724 $\pm$ 0.011 & 0.537 $\pm$ 0.005 & 1.262 $\pm$ 0.010 \\ \cline{2-11}
 & \multirow{ 3}{*}{Stage 2} & ATST  & 1e-5 & 1e-4 & 1e-4 & 0.5 & avg. pool & 0.75 $\pm$ 0.004 & 0.548 $\pm$ 0.004 & \textbf{1.298} $\pm$ 0.006  \\
&  & fPaSST & 5e-5 & 5e-4 & 1e-4 & 1 & int. nearest & 0.719 $\pm$ 0.013 & 0.539 $\pm$ 0.003 & 1.259 $\pm$ 0.015  \\
&  & BEATs & 5e-5 & 5e-4 & 1e-4 & 1 & int. nearest & 0.7286 $\pm$ 0.005 & 0.557 $\pm$ 0.005 & 1.286 $\pm$ 0.009 \\ 
\bottomrule
\end{tabular}
\caption{
 The table presents the results of ATST, fPaSST, and BEATs for both iterations and stages. For each model, we list the best configuration in terms of the sequence length adaptation method (\textbf{Seq.}). \textit{Ensemble Stage 2} is used to generate the pseudo-labels for Iteration 2. \textbf{Rank Score} denotes the sum of \textbf{mpAUC} and \textbf{PSDS1}.
}
 \label{tab:results}
 \vspace{-3 pt}
\end{table*}

\begin{table*}[t]
\begin{center}
\begin{tabular}{@{}cccc|cccc@{}}
\toprule
ID & \textbf{Models} & \textbf{\#} & \textbf{Dev-Test} & \textbf{mpAUC} & \textbf{PSDS1 MF} & \textbf{PSDS1 SEBB} & \textbf{Eval PSDS1 SEBB}  \\ \midrule
\textbf{S1} & ATST I2.S2 & 1 & \xmark & 0.749 & 0.548 & 0.617 & 0.684 \\ \midrule
\textbf{S2} & ATST I2.S2 & 1 & \cmark & - & - & - & 0.692 \\  \midrule
\textbf{S3} & Ensemble I2.S1 + I2.S2 & 18 & \xmark &  0.743 & 0.569 & 0.632 & 0.721 \\  \midrule
\textbf{S4} & Ensemble I2.S1 + I2.S2 & 15 & \cmark & - & - & - & 0.729 \\ 
\bottomrule
\end{tabular}
\caption{Final Submissions: \textbf{\#} lists the number of models; the flag \textbf{Dev-Test} indicates that we use the full development set for training; \textbf{PSDS1 MF} lists results with a median filter; \textbf{PSDS1 SEBB} lists DESED test set results using SEBB postprocessing~\cite{ebbers2024sound}; and \textbf{Eval PSDS1 SEBB} lists results on the public evaluation set with SEBB postprocessing. }
\label{tab:sub}
\end{center}
\vspace{-16pt}
\end{table*}

\vspace{-5pt}
\subsection{Postprocessing}
\label{subsec:post}

For model selection and hyperparameter tuning, we stick with the class-wise median filter used in the baseline system~\cite{cornell2024dcase}. After selecting models for submission, we apply the recently introduced Sound Event Bounding Boxes~\cite{ebbers2024sound} method for post-processing. We use class-wise parameters and obtain them by using linearly spaced search grids (8 values) for step filter length (0.38 to 0.66), relative merge threshold (1.5 to 3.25), and absolute merge threshold (0.15 to 0.325). We follow the strategy of the baseline~\cite{cornell2024dcase} and tune these hyperparameters on the development-test set, as the class-wise median filter lengths of the baseline system are tuned on the development-test set as well.

\vspace{-5pt}
\section{Results}
\label{sec:results}

In this section, we present the results of the described models (Section~\ref{sec:arch}) in the introduced training pipeline (Section~\ref{sec:training_setup}). In Section~\ref{subsec:final_sub}, systems selected for submission are presented.

Table~\ref{tab:results} lists the results for the best configuration of each model in terms of sequence length adaptation method (\textbf{Seq.}) and learning rate in each iteration and stage. Furthermore, the CNN (\textbf{lr\_cnn}), RNN (\textbf{lr\_rnn}), and Transformer (\textbf{lr\_tf}) learning rates are listed. \textbf{lr\_dec} describes layer-wise learning rate decay for the Transformer models as used in~\cite{shao2024fine}. 

In I1.S1, in which the transformer models are frozen, BEATs seems to extract the embeddings of the highest quality, followed by fPaSST and ATST. I1.S1 with BEATs is very similar to the baseline setup~\cite{cornell2024dcase} and achieves a similar rank score with a slight performance increase in our setup.

Compared to the rank scores in I1.S1, in I1.S2, all three transformers demonstrate a large increase in rank score when fine-tuned on the Task 4 datasets. Notably, the three transformer models have different strengths, with ATST achieving the best score on MAESTRO clips while BEATs obtains the best score on DESED clips. \textit{Ensemble Stage 2} denotes an ensemble of 46 models resulting from I1.S2, including ATST, fPaSST, and BEATs trained in different configurations. While the performance in terms of PSDS1 benefits largely from ensembling a large number of models, the mpAUC is even slightly worse compared to the best single model after I1.S2 (ATST). We use \textit{Ensemble Stage 2} to generate strong pseudo-labels for all audio clips in the dataset.

The additional pseudo-label loss in I2.S1 boosts performance substantially, with all three transformers achieving higher performance in I2.S1 in terms of rank score compared to I1.S2. Interestingly, ATST, which achieves the lowest performance in I1.S1, has the highest performance in I2.S1 outperforming the other models in particular on the MAESTRO clips. 

The top rank scores for all models are achieved in I2.S2, with ATST obtaining the best single-model performance. Notably, the pseudo-label loss is not used in I2.S2, as it does not increase the rank score, demonstrating that a well-trained CRNN from S1 is instrumental for high performance in S2.

\vspace{-5pt}
\subsection{Final Submissions}
\label{subsec:final_sub}

For the final submissions shown in Table~\ref{tab:sub}, we select the top single-model, ATST, after I2.S2, and build an ensemble consisting of ATST, fPaSST, and BEATs models obtained in I2. We repeat the full training process for the two selections and include the test data of MAESTRO and DESED for training to make use of the full development set. In this case, model selection still relies on validation metrics.

As described in Section~\ref{subsec:post}, we use SEBBs~\cite{ebbers2024sound} instead of a class-wise median filter for postprocessing the predictions of all submissions. The resulting performance is listed in Table~\ref{tab:sub}. \textbf{PSDS1 MF} and \textbf{PSDS1 SEBB} denote the performance on the DESED test set with median filter and SEBB, respectively. Although the median filter lengths of the baseline system are also tuned on the test set, we note that \textbf{PSDS1 SEBB} results should be taken with a grain of salt, as the SEBB hyparameters are tuned on the test set. We therefore also report the results on the unseen DESED public evaluation set (\textbf{Eval PSDS1 SEBB}). Notably, our best single model (\textbf{S2}) improves the state-of-the-art PSDS1 score on the public evaluation set from 0.686~\cite{ebbers2024sound} to 0.692. The submitted ensembles (\textbf{S3} \& \textbf{S4}) clearly improve over the single models (\textbf{S1} \& \textbf{S2}) in terms of PSDS1, but interestingly, mpAUC cannot be improved by ensembling. 

\vspace{-5pt}
\section{ACKNOWLEDGMENT}
\label{sec:ack}

The computational results presented were achieved in part using the Vienna Scientific Cluster (VSC) and the Linz Institute of Technology (LIT) AI Lab Cluster. The LIT AI Lab is supported by the Federal State of Upper Austria. Gerhard Widmer's work is supported by the European Research Council (ERC) under the European Union's Horizon 2020 research and innovation programme, grant agreement No 101019375 (Whither Music?).

\bibliographystyle{IEEEtran}
\bibliography{refs}
%
%
%
%
%
%
%
%
%

\end{sloppy}
\end{document}